\documentstyle[psfig]{article}

\def\lcdm{$\Lambda$CDM}
\def\gsim{\mathrel{\raise.3ex\hbox{$>$\kern-.75em\lower1ex\hbox{$\sim$}}}}
\def\lsim{\mathrel{\raise.3ex\hbox{$<$\kern-.75em\lower1ex\hbox{$\sim$}}}}

\def\M10{{\times 10^{10} M_{\odot}\ }}

\def\kmsMpc{\ {\rm km}\ {\rm s}^{-1}\ {\rm Mpc}^{-1}}

\begin{document}

\title{Favored Variants of Cold Dark Matter Cosmologies
\footnote{To appear in the proceedings of ``Identification of Dark Matter'',
University of Sheffield, UK (World Scientific, 1996)}
}

\author{Rachel S. Somerville and Joel R. Primack \\
Physics Department, University of California,
Santa Cruz CA 95064\\
rachel@physics.ucsc.edu; joel@ucolick.org}

\maketitle

\begin{abstract}
We discuss variants of Cold Dark Matter (CDM) that give good agreement
with a range of observations. We consider models with hot dark matter,
tilt, $\Omega < 1$, or a cosmological constant. We also discuss the
sensitivity of the results to other parameters, such as the Hubble
parameter and the baryon fraction. We obtain constraints by combining
the COBE data, cluster abundances, abundance of damped Lyman-$\alpha$
systems at $z\sim3$, the small-angle Cosmic Microwave Background
anisotropy, and the small-scale non-linear power spectrum. We present
non-linear power spectra from a new suite of N-body simulations for
the ``best-bet'' models from each category.

\end{abstract}

\section{Introduction}

Since CDM comes close to explaining many fundamental observations, it
has now become popular to consider models that are mostly CDM, with a
little bit of something else.  There is no shortage of possibilities
for the ``something else,'' and at present neither the total mass
density $\Omega_0$, the baryonic density $\Omega_b$, the Hubble
parameter $H_0 = 100 h \kmsMpc$, nor the power-law index $n$ of the
spectrum of primordial fluctuations ($P(k) \propto k^n$) are very well
constrained.  However, it is an interesting exercise to see what
corner of parameter space large scale structure observations alone
would push us into, for various classes of models. This will be the
primary focus of this discussion. We will present ``best'' versions of
the four most popular classes of CDM variants, based on a comparison
of semi-analytic calculations and N-body simulations with various
observations. The four classes of variants considered are Cold plus
Hot Dark Matter (CHDM)\cite{PHKC95} and tilted CDM
(TCDM)\cite{WVLS96}, both with $\Omega=1$, and low-$\Omega$ CDM models
in either a flat cosmology with a cosmological constant $\Lambda$ such
that $\Omega_\Lambda \equiv \Lambda/(3H_0^2) = 1 - \Omega_0$
(\lcdm)\cite{LLRV96} or an open cosmology with $\Lambda=0$
(OCDM).\cite{LLVW96} As this exercise is to some degree a matter of
taste, we will discuss the reasoning behind our choice of each
parameter set. For more on the current observational constraints on
the fundamental cosmological parameters, see e.g. Primack.\cite{P96}

\section{Constraints from Large Scale Structure}
\subsection{Cluster Abundance}
\label{sec:clusters}
The abundance of clusters $N(>M)$ combined with the COBE normalization
provides a strong constraint on cosmological models.  The number
densities of clusters with X-ray temperatures exceeding $k_BT = 3.7$
keV and 7 keV were measured by Henry \& Arnaud.\cite{HA91}  The main
uncertainty in $N(>M)$ lies in the translation from velocity
dispersion or X-ray luminosity to mass.  The X-ray mass estimates
could be affected by temperature gradients, substructure, or
ellipticity.\cite{tkb94} We have taken the mass range for the $k_BT >
3$keV clusters from White et al.\cite{wef93} and the mass range for
the $k_BT > 7$ keV clusters from Liddle et al.\cite{LLRV96}

\begin{figure}[!htbp]
%\centerline{\epsfxsize\columnwidth\epsfysize\columnwidth\epsfbox{nclust.ps}}
\centering
\centerline{\psfig{file=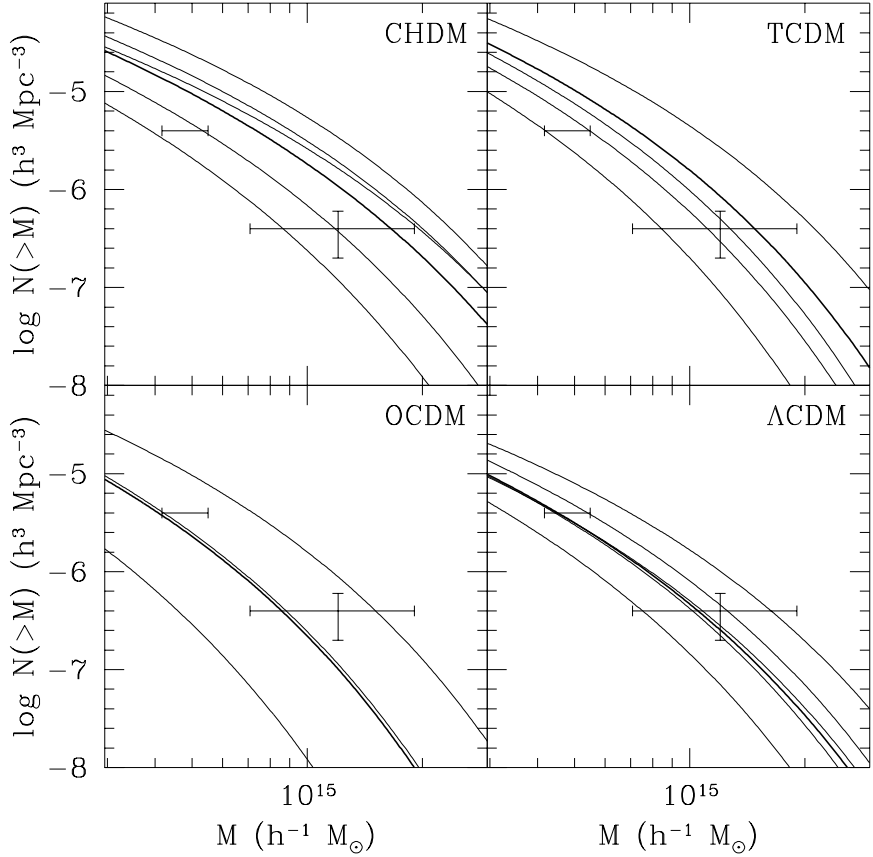,width=\columnwidth}}
\caption{Number density of halos with mass greater than $M$ from a
Press-Schechter approximation with $\delta_c=1.69$ and a top-hat
filter. Points show the number density of clusters with X-ray
temperature exceeding 3 keV (upper left) and 7 keV (lower right)
\protect\cite{HA91}. In order of decreasing number density, the models
shown are: CHDM $N_\nu=1$ and $\Omega_\nu=0.1$, 0.2, and 0.3;
$\boldmath{N_\nu=2}$ and $\boldmath{\Omega_\nu=0.2}$; $N_\nu=1$,
$\Omega_\nu=0.2$, and $n=0.95$; and $N_\nu=1$, $\Omega_\nu=0.2$, and
$n=0.9$. TCDM: $n=0.9$ and $h=0.5$; $\boldmath{n=0.9}$ and
$\boldmath{h=0.45}$; $n=0.8$ and $h=0.5$; $n=0.9$ and $h=0.42$;
$n=0.8$ and $h=0.45$. OCDM: $\Omega_0=0.7$ and $h=0.5$; $\Omega_0=0.6$
and $h=0.5$; $\boldmath{\Omega_0=0.5}$ and $\boldmath{h=0.6}$;
$\Omega_0=0.4$ and $h=0.65$. $\Lambda$CDM: $\Omega_0=0.4$, $h=0.65$;
$\Omega_0=0.4$, $h=0.60$; $\Omega_0=0.3$, $h=0.70$;
$\boldmath{\Omega_0=0.4}$, $\boldmath{h=0.60}$, $\boldmath{n=0.9}$,
$\boldmath{\sigma_8=0.88}$; $\Omega_0=0.4$, $h=0.65$, $n=0.9$;
$\Omega_0=0.4$, $h=0.60$, $n=0.9$. Bold lines indicate the ``favored''
models discussed in the text. Parameters not specified are as given in
Table 1.}
\label{fig:nclust}
\end{figure}

Figure \ref{fig:nclust} shows the cluster abundances obtained from the
Press-Schechter approximation\cite{ps74} for a variety of models (see
the figure caption). The CHDM model with $\Omega_\nu= 0.2$ in one
species of massive neutrinos ($N_\nu=1$) overproduces clusters by a
factor of about two.  The lower neutrino mass that arises with the
same $\Omega_\nu$ but $N_\nu =2$ species of neutrinos results in a
longer free-streaming length, which decreases the power by about 20\%
on cluster scales without significantly affecting larger or smaller
scales.\cite{PHKC95,PogS} The CHDM model with $N_\nu=2$ is in better
agreement, but still too high on the basis of this analysis. However,
moderate over-prediction of the cluster abundance is probably not too
worrisome, since cluster masses derived from gravitational lensing
tend to be systematically higher than the X-ray masses by as much as a
factor of two.\cite{clustermass} Also, preliminary work with
simulations suggests that the Press-Schechter approximation is
systematically high on these scales.\cite{gsphk96} Introducing a very
moderate tilt ($n=0.95$) would bring this model into agreement with
the cluster data in Figure \ref{fig:nclust}, but this would reduce the
power on small scales, leading to less early structure formation.

For the TCDM models, the best agreement is obtained with either
$n=0.8$ and $h=0.5$ or $n=0.9$ and $h\sim0.45$. For OCDM, one can see
that there is no hope of producing enough clusters with $\Omega <
0.5$. The only way to improve the very low-$\Omega$ models would be to
introduce a ``positive tilt'' ($n>1$). One can see that there is an
almost perfect degeneracy in $N(>M)$ between $\Omega_0$ and $h$.
$\Lambda$CDM models with $\Omega_0 = 0.3$ or $\Omega_0=0.4$ with some
tilt ($n=0.9$) agree reasonably well with the cluster data.

\subsection{Early Structure Formation}
\begin{figure}[!htb]
%\centerline{\epsfxsize=8truecm\epsfysize=8truecm\epsfbox{ogasz.ps}}
\centering
\centerline{\psfig{file=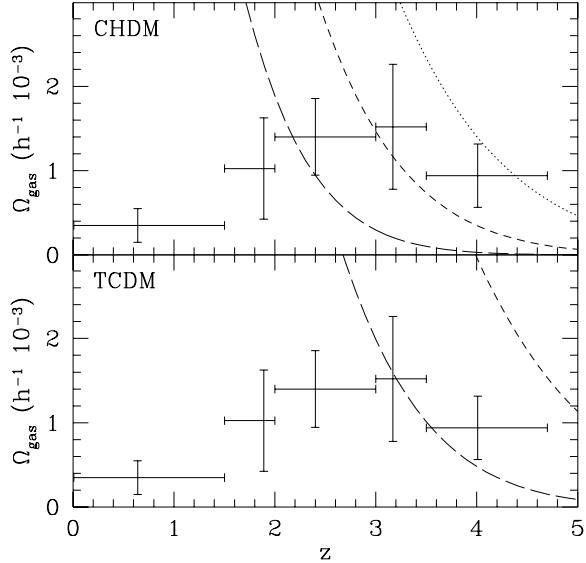,width=8truecm}}
\caption{Collapsed gas as a function of redshift, from a
Press-Schechter analysis with $\delta_c=1.4$ and top-hat
smoothing (cf. Klypin et al.\protect\cite{KBHP95} and references therein).
Long dashed, short dashed, and dotted lines correspond to
halos with masses of $10^{12}$, $10^{11}$, and $10^{10}
h^{-1}M_{\odot}$ respectively. Data points are from Storrie-Lombardi 
et al.\protect\cite{sl96} Models shown are CHDM $\Omega_\nu=0.2$, $n=1$,
$N_\nu=1$ and TCDM $n=0.9$, $h=0.45$.}
\label{fig:dlasz}
\end{figure}

Observations of quasar absorption systems (clumps of gas or
proto-galaxies that produce absorption features in quasar spectra)
place constraints on structure formation up to $z \sim4$. We will
focus on observations of high-density absorption systems called damped
Lyman-$\alpha$ systems (DLAS). The Press-Schechter approximation can
again be used to provide a lower limit on the abundance of collapsed
gas at a given redshift --- see Figure~\ref{fig:dlasz}. Some of the
gas may be ionized or consumed by star formation, so a viable model
should predict at least as much collapsed gas as the observations. A
more detailed analysis including physical modeling of gas cooling,
star formation, and supernova feedback is necessary in order to
calculate the actual amount of gas in absorption systems.\cite{spf96}
It should be noted that the measurement of $\Omega_{gas}$ at
$z=$3--3.5 has come down by nearly a factor of three due to new
observations,\cite{sl96} easing the constraint on CHDM models. The new
observations also suggest that $\Omega_{gas}$ may peak at $z\sim3$ and
fall off at higher redshifts, which could be explained very naturally
by the decreased supply of collapsed gas predicted in CHDM models.

If the DLAS have relatively small masses ($\sim 10^{10}$--$10^{11}
M_\odot$), then CHDM models with $\Omega_\nu \sim 0.2$ produce enough
collapsed gas to be compatible with these observations, even with a
small tilt ($n=0.95$). Models with higher $\Omega_\nu$ or larger tilt
probably will not produce enough collapsed gas. The models with
$N_\nu=2$ would produce very similar results because changing the
number of neutrino species does not affect the power on small scales.
Figure~\ref{fig:dlasz} (lower panel) shows the predicted
$\Omega_{gas}$ for tilted models with pure CDM. These models have no
difficulty producing enough collapsed gas. The OCDM and LCDM models
have even earlier structure formation and produce plenty of collapsed
gas at high redshift --- but they might not easily account for a
fall-off of $\Omega_{gas}$ with $z$.

Recent observations of ``normal'' (i.e. non-AGN) galaxies at high
redshift \cite{s96} may provide another constraint on early structure
formation. The masses of these objects are uncertain, however, so a
comparison with theory requires a fairly detailed treatment of galaxy
formation~\cite{sfp96}.

\section{Summary of ``Best'' CDM Variants}

Taking all the constraints into account, we can arrive at ``best''
choices for each of the classes of models.  The parameters for these
models are summarized in Table 1. We have run a suite of large,
high-resolution Particle Mesh (PM) N-body simulations
\cite{gsphk96} of these models, in order to study their
non-linear clustering and large scale structure properties in more
detail.  We will now summarize the considerations that led to our
choices of parameters for each model.

%model parameters table
\begin{table*}[!htb]
\begin{tabular}{llllllllll}
%\tableline
\noalign{\hrule}
Model& $t_0$& $h$ & $\Omega_0$ & $\Omega_b$& $\Omega_{\nu}$& 
  $\Omega_\Lambda$ & $n$ & $N_\nu$& $\sigma_8$\\
%\tableline
\noalign{\hrule}
SCDM         &13.0&0.50 &1.0   &0.0   &0.0 &0.0 &1.0 &0&0.667\\
TCDM         &14.5&0.45&1.0  &0.1  &0.0 &0.0 &0.9&0&0.732\\
CHDM-2$\nu$  &13.0&0.50 &1.0  &0.1  &0.2&0.0 &1.0 &2&0.719\\
tCHDM-2$\nu$ &10.9&0.60 &1.0&0.069&0.2&0.0 &0.9&2&0.677\\
OCDM         &14.7&0.60 &0.5&0.069&0.0 &0.0 &1.0 &0&0.773\\
t$\Lambda$CDM&14.5&0.60 &0.4&0.035&0.0 &0.6&0.9&0&0.878\\
%\tableline
\noalign{\hrule}
\end{tabular}
\caption{Model Parameters}
\end{table*}

\subsection{CHDM}

We chose two CHDM models to investigate. The $\Omega_\nu =0.2$ and
$N_\nu = 2$ model with $n=1$ gives reasonable agreement with the
cluster data (see the caveat in section \ref{sec:clusters}) and has
ample collapsed gas at $z\sim3$ to be compatible with the DLAS data.
We also consider a tilted CHDM model with a high Hubble parameter ($h
= 0.6$).  The tilt reduces the power on small scales, but the
increased Hubble parameter partially compensates for this, so this
model still gives good agreement with both the cluster data and the
DLAS data. However, the age of the universe is only 11 Gyr, several
Gyr younger than most estimates of globular cluster ages. But if the
evidence for a large Hubble parameter becomes more certain, this model
might be worth taking seriously.

\subsection{TCDM}

We chose to investigate a tilted model with $n=0.9$ and $h=0.45$. A
model with $n=0.8$ and $h=0.5$ gives very similar results for the
cluster abundance, but is disfavored by the Saskatoon data on the CMB
anisotropy at small angular scales.\cite{CMB} As for the CHDM models,
we have chosen a rather high $\Omega_b$, consistent with the low D/H
measured by Tytler et al.,\cite{BT96} to lessen the conflict with
cluster baryons for these $\Omega=1$ models.\cite{clusterbaryon}

\subsection{OCDM}

Several independent analyses of galaxy peculiar velocities lead to
strong lower limits on $\Omega_0$ in all models with Gaussian
primordial fluctuations.\cite{Dekel} It is interesting that one would
reach similar conclusions based only on the cluster data --- open
models with $\Omega_0 \leq 0.4$ are very strongly ruled out because
they simply do not produce nearly enough clusters. Even open models
with $\Omega_0 = 0.5-0.6$ will probably underproduce clusters. We have
chosen a model with $\Omega_0=0.5$ and $h=0.6$. This model is almost
completely degenerate with a model with higher $\Omega_0$ and lower
$h$. We chose this model because of the interest in models with higher
values of the Hubble Parameter.

\subsection{$\Lambda$CDM}

Previously favored $\Lambda$ models with $\Omega_0=0.3$,
$\Omega_\Lambda=0.7$, and $h=0.7$ give good agreement with cluster
abundance data but produce far too much power on small
scales.\cite{kph} This is difficult to reconcile with physical models
of galaxy formation, which predict that the galaxy power spectrum
should if anything be larger than the matter power
spectrum.\cite{kns95} If one is unwilling to accept a scale-dependent
anti-bias of galaxies with respect to dark matter, the only way to
save this class of models is to go to higher $\Omega_0$ (thus smaller
$\Omega_\Lambda$, also favored by recent data on quasar
lensing\cite{Kochanek}, HST galaxy counts\cite{Driver}, and $z \sim
0.5$ supernovae\cite{Perl96}), smaller $h$, and to add a tilt to
reduce the small scale power. We have chosen a model with $\Omega_0 =
0.4$, $h=0.60$, and a tilt of $n=0.9$. As we saw in
Figure~\ref{fig:nclust}, when normalized to the central COBE value
this model underproduces clusters. We have therefore normalized the
model to an amplitude about 10\% higher than the central value, but
still within the 1$\sigma$ uncertainty. The same model normalized to
the central COBE value but with a slightly higher Hubble parameter
$h=0.65$ would also fit the cluster data, but may have too much power
on small scales.

\subsection{Non-linear Power Spectra}

\begin{figure}[!htb]
%\centerline{\epsfxsize\columnwidth\epsfbox{power.ps}}
\centering
\centerline{\psfig{file=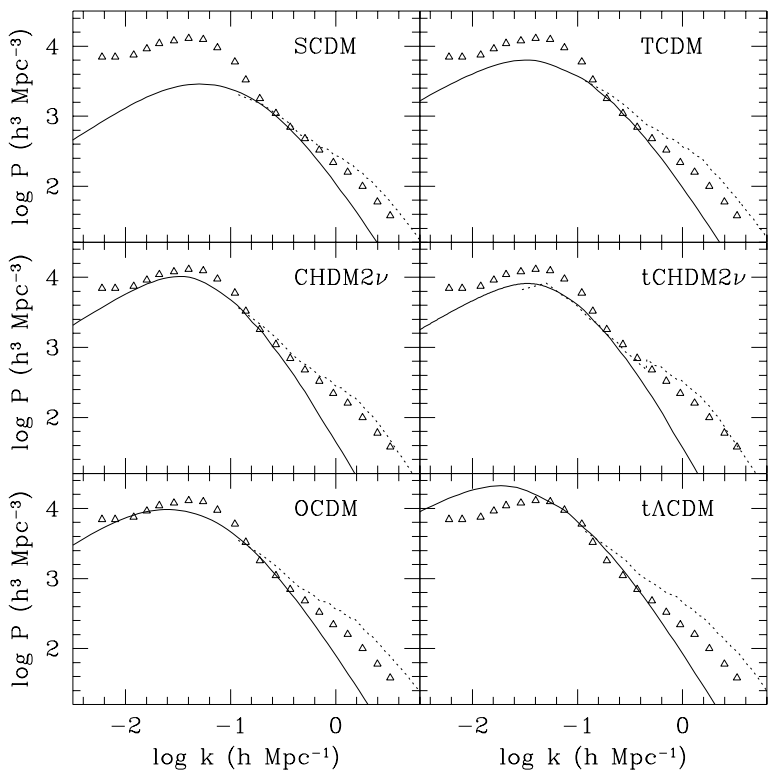,width=11cm}}
\caption{Power Spectra. Solid lines are the linear power and dotted
lines are the non-linear power from N-body simulations
\protect\cite{gsphk96}. Triangles are the galaxy power spectrum from
the APM redshift survey \protect\cite{apm}. Model parameters are given
in Table~1. }
\label{fig:power}
\end{figure}

Figure~\ref{fig:power} shows the linear power spectra for our
``favorite'' models and the non-linear power spectra from the N-body
simulations. All of the models appear to agree reasonably well with
the APM real space power spectrum (plotted with triangles), however
one should keep in mind that there is still the issue of galaxy bias
to contend with. The OCDM and t$\Lambda$CDM models in particular may
still have too much power on small scales. This needs to be
investigated using a physical model for galaxy bias,\cite{sgp96} which
will also permit investigation of many other potentially powerful
statistics of the galaxy distribution on nonlinear scales, such as the
pairwise velocity,\cite{RPD} group velocity dispersion,\cite{NKP}
filament statistics,\cite{Dave} or the void probability
function.\cite{Ghigna}

\section{Conclusions}

Although we have argued that the parameters summarized in Table 1 are
the ``best'' choices for these classes of models, each of these models
is potentially inconsistent with at least one observation. The CHDM
models that we have discussed may have great difficulty explaining
high redshift galaxies and other data on structure formation at high
redshift.\cite{sfp96} In addition, any $\Omega=1$ model may be
difficult to reconcile with measurements of the baryon fraction in
clusters,\cite{clusterbaryon} and will correspond to a worrisomely
young universe if the Hubble parameter is large. The TCDM model has $h
< 0.5$ and is barely consistent with the small-angle CMB data. The
OCDM and $\Lambda$CDM models may produce too much power on galaxy
scales relative to cluster scales.

Our knowledge of the fundamental cosmological parameters is likely to
improve dramatically in the next decade with the launch of the next
generation of CMB satellites (MAP and COBRAS/SAMBA). In the meantime,
a ``phenomenological'' approach to cosmology may enable us to make
progress towards understanding galaxy formation and evolution as well
as large scale structure, all of which may remain difficult problems
even after the cosmological parameters are determined. In addition,
the efforts to identify and directly detect the dark matter are
extremely important in the effort to form a firmer theoretical
foundation for Cold Dark Matter dominated cosmology.

\section*{Acknowledgements}
R.S.S. was supported by a GAANN fellowship, and J.R.P. was supported
by NASA and NSF research grants, at UCSC.  The simulations were done
on the SP2 supercomputer at the Cornell Theory Center.

%\section*{References}

\clearpage

\end{document}